\documentclass[12pt]{iopart}

\usepackage{color}
\usepackage{url}
\usepackage[dvipdfm]{graphicx}
\usepackage{hangcaption}

\newcommand{\fft}[1]{FFT \biggl[#1 \biggl]} 
\newcommand{\ifft}[1]{FFT^{-1} \biggl[#1 \biggl]} 

\begin{document}

\title{Real-time digital holographic microscopy observable in multi-view and multi-resolution}

\author{Tomoyoshi Shimobaba, Nobuyuki Masuda, Yasuyuki Ichihashi, and Tomoyoshi Ito}

\address{Graduate school of engineering, Chiba University, 1-33 Yayoi-cho, Inage-Ward, Chiba 263-8522, Japan}

\ead{shimobaba@faculty.chiba-u.jp}
\begin{abstract}

We propose a real-time digital holographic microscopy, that enables simultaneous multiple  reconstructed images with arbitrary resolution, depth and positions, using Shifted-Fresnel diffraction instead of Fresnel diffraction. 
In this system, we used four graphics processing units (GPU) for multiple reconstructions in real-time.
We show the demonstration of four reconstruction images from a hologram with arbitrary depths, positions, and resolutions.

\end{abstract}


\maketitle

\section{Introduction}

 In research fields such as Micro Electro Mechanical Systems (MEMS) and bio-imaging, digital holographic microscopy (DHM) \cite{dh,poon,takaki} is attractive as a new microscopy technique, because the DHM allows both the amplitude and phase of a specimen to be simultaneously observed. 

The technique can obtain a hologram whereby the information of a specimen is electronically recorded, via the use of a charge-coupled device image sensor (CCD). 
In order to obtain a reconstructed image from a hologram, numerous calculations for the Fresnel diffraction \cite{ersoy,goodman} are required; however, the Fresnel diffraction has a restriction, namely, the same sampling spacings must be set on the hologram and the reconstructed image, or the sampling spacing on the reconstructed plane depends on the propagation distance and the wavelength of a reference light. 
Therefore, we cannot observe the reconstructed image with arbitrary resolution due to the restriction of the sampling spacing. 
In addition, in current DHM, an objective lens is used to increase the resolution of the reconstructed image; however, using the objective lens sacrifices an area of the reconstructed image. 

In this paper, without using an objective lens, we propose a DHM observable in multi-view and multi-resolution. 
The DHM can obtain multiple reconstructed images with arbitrary resolution, depths and positions, using Shifted-Fresnel diffraction \cite{shift}, instead of Fresnel diffraction. 
Shifted-Fresnel diffraction based on Fresnel diffraction can calculate a reconstructed image with different sampling spacings between the hologram and the reconstructed image, as well as a shift away from the propagation axis. 
In addition, we used four graphics processing units (GPU) chips \cite{gpugems} in order to observe four reconstructed images in real-time from one hologram.

In Section 2, we describe the concept and the calculation method for the proposed DHM. 
In Section 3, we describe the results of an optical experiment. 
In Section 4, we conclude this work.
In Section 4, we conclude this work.

\section{Digital holographic microscopy using the Shifted-Fresnel diffraction}

\begin{figure}[htb]
\centerline{
\includegraphics[width=7cm]{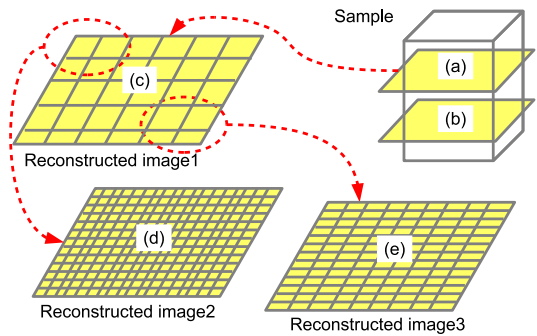}}
 \caption{ The concept of the proposed DHM.}
\label{fig:system}
\end{figure}

\begin{figure}[htb]
\centerline{
\includegraphics[width=7cm]{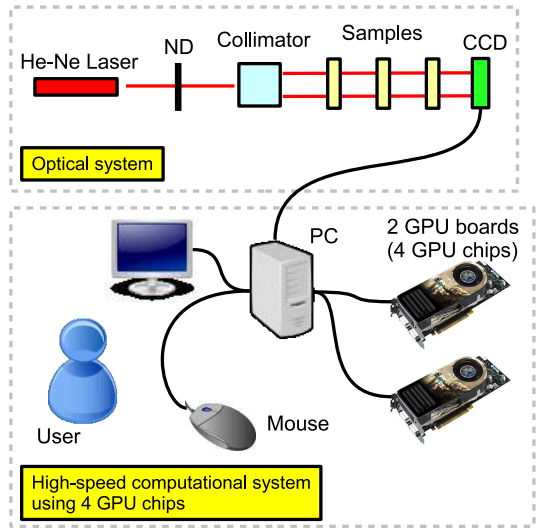}}
 \caption{The outline of the proposed DHM system.}
\label{fig:system-photo}
\end{figure}

The concept of the proposed DHM is shown in Fig.1. 
In the DHM, we can simultaneously observe reconstructed images at different depths (Fig.1 (a) and (b)) along the depth direction. 
In addition, while observing a wide viewing area of a reconstructed image (Fig.1(c)), we can simultaneously observe reconstructed images with the user-setting arbitrary resolutions at arbitrary positions (Fig.1 (d) and (e)).

An objective lens can increase the resolution of the reconstructed image, while decreasing the viewing area of the reconstructed image.
For this reason, the proposed DHM system does not use an objective lens, instead, the proposed DHM system uses arbitrary sampling spacing on the reconstructed image.
If we observe a wide viewing area of a reconstructed image, we set a large sampling spacing on the reconstructed plane.
Therefore in order to observe the reconstructed image in detail, 
we need to set a small sampling spacing on the reconstructed plane.

In order to realize the proposed DHM, the following methods are required: 
\begin{enumerate}
\item A computational method for obtaining reconstructed images with arbitrary resolution from a hologram 
\item A high-speed computational system for reconstructing from a hologram in real-time. 
\end{enumerate}

To solve these problems, we used Shifted-Fresnel diffraction and multi GPUs.
An outline of the DHM system is shown in Fig.2. 
The system consists of an optical system without using an objective lens, and a high-speed computational system using four GPU chips.
Some researchers have already used the GPU approach for real-time digital holographic reconstruction \cite{realdhm, dhgpu,dhgpu2}. 
In the figure, we used a 5-mW He-Ne laser (the wavelength is 632.8 nm) as a reference light. 
"ND" indicates a neutral density filter. 
We used a CCD camera, which has a resolution of $1,360 \times 1,024$ and a pixel pitch of $4.65 \mu m \times 4.65 \mu m$. 
In the reconstruction calculation from a hologram captured by the CCD, we resize the hologram with $1,024 \times 1,024$ grids in order to use fast Fourier transform (FFT) for the calculation. 
We also used three samples, which are USAF 1951 test target, the head of a mosquito and a fly.
Holograms captured by the CCD are recorded as in-line Gabor hologram and are transferred to personal computer via the USB2.0 interface.
Then, four GPU chips calculate four reconstructed images from one hologram in real-time.

 \subsection{Shifted-Fresnel diffraction}

Recently, a new diffraction calculation, Shifted-Fresnel diffraction, has been proposed \cite{shift}. 
The method enables arbitrary sampling spacings to be set on a hologram and a reconstructed plane as well as a shift away from the propagation axis. 
Other methods capable of changing sampling spacing have also been studied  \cite{resolution1,resolution2}.
We chose Shifted-Fresnel diffraction because one can do a shift away from the propagation axis.
 Shifted-Fresnel diffraction is expressed by the following equations:

\begin{eqnarray}
u_1[m_1,n_1] &=&
C_1 \sum_{m_0} \sum_{n_0} u_0'[m_0,n_0] h[m_1-m_0, n_1-n_0]\nonumber \\ 
&=& C_1 u_0'[m_1,n_1] * h[m_1, n_1] \nonumber \\ 
&=& C_1 \ifft{ \fft{u_0'[m_1,n_1]} \fft{h[m_1, n_1]} } 
\end{eqnarray}
\begin{eqnarray}
C_1&=& \frac{\exp(i k z)}{i \lambda z} \exp(i \frac{\pi}{\lambda z} (x_1^2+y_1^2))  
\exp(-i \frac{2 \pi}{\lambda z}(( - \frac{N_x}{2} + O_{x_0}) p_{x_0} x_1 +  \nonumber \\
& & ( - \frac{N_y}{2} + O_{y_0}) p_{y_0} y_1  ))  \exp(- i \pi(S_x m_1^2 + S_y n_1^2))
\end{eqnarray}
\begin{eqnarray}
u_0'[m_0,n_0]& =& u_0[m_0,n_0] \exp(i \frac{\pi}{\lambda z}(x_0^2+y_0^2))
			    \exp(-i 2 \pi (m_0 S_x( - \frac{N_x}{2} + O_{x_1})  +  \nonumber \\ 
			& & n_0 S_y( - \frac{N_y}{2} + O_{y_1}) )) 
			    \exp(- i \pi (S_x m_0^2 + S_y n_0^2 ))
\end{eqnarray}
\begin{eqnarray}
h[m_1-m_0,n_1-n_0] &=& \exp(i \pi( S_x(m_1-m_0)^2 + S_y(n_1-n_0)^2 )) 
\end{eqnarray}
where, the operators $FFT$ and $FFT^{-1}$ denote the FFT and the inverse FFT, $i$ is $\sqrt{-1}$, $\lambda$is the wave length of the reference light, $z$ is the distance between the hologram and the reconstructed image, $[m_0,n_0]$ and $[m_1, n_1]$are the discretized coordinates on the hologram and the reconstructed image, $u_0[m_0,n_0]$ and $u_1[m_1,n_1]$ are the hologram and the reconstructed image, $p_{x_0}$ and $p_{y_0}$ are the sampling spacing on the hologram,  $p_{x_1}$ and  $p_{y_1}$ are the sampling spacing on the reconstructed image,  $(O_{x_0}$,  $O_{y_0})$ and $(O_{x_1}$,  $O_{y_1})$ are the shift distances away from the propagation axis. 
And, we define 
$S_x=\frac{p_{x_0} p_{x_1}}{\lambda z}$ , $S_y=\frac{p_{y_0} p_{y_1}}{\lambda z}$, 
$x_0=m_0 p_{x_0} + O_{x_0}$, 
$y_0=n_0 p_{y_0} + O_{y_0}$, 
$x_1=m_1 p_{x_1} + O_{x_1}$, and  
$y_1=n_1 p_{y_1} + O_{y_1}$. 
For more details, see Ref.\cite{shift}.

Last, we can obtain the light intensity as a reconstructed image using the following equation:
\begin{eqnarray}
I[m_1,n_1]=|u_1[m_1,n_1]|^2
\end{eqnarray}

Calculating Eq.(1) and Eq.(5), we can obtain a reconstructed image with an arbitrary depth, resolution and shift to change the parameters $z$, $p_{x_1}, p_{y_1}$ and $O_{x_1}, O_{y_1}$, respectively. 
Note that we can neglect the coefficient $C_1$  because we need the light intensity.

\subsection{Fast calculation of the Shifted-Fresnel diffraction using multi-GPU}

Recent GPUs with many stream processors allow us to use highly parallel processors. 
The stream processor can operate 32-bit (or 64-bit) floating-point addition, multiplication, and multiply-add instructions. 
We have already reported a real-time DHM system based on the Fresnel diffraction using a GPU \cite{realdhm}. 

In this paper, for the Shifted-Fresnel diffraction, we used the GPU-based Wave Optics (GWO) library \cite{gwo}. 
The library is a numerical calculation library for the diffraction calculations using a GPU. 
If optics engineers and researchers have no knowledge of GPU, the GWO library provides them with the GPU computation power easily. 
The current GWO library runs on Microsoft Windows XP and is provided as a dynamic link library (DLL). 
The library and a sample code can be downloaded from Ref.\cite{gwourl}.

As shown in Eq.(1), Shifted-Fresnel diffraction can accelerate the computational time using two FFTs and one inverse FFT; however, recent central processing units (CPUs) do not have sufficient computational power for real-time calculation. 
Therefore, we use GPU instead of CPU .

Multiple reconstructions using multi-GPU is shown below:

\begin{enumerate}
\item The CPU send a hologram captured by the CCD to the memories on the two GPU boards.
\item Each GPU chip expands the doubled size of the hologram ($2,048 \times 2,048$  grids) with zero-padding to avoid the circular convolution.
\item Each GPU chip calculates the complex multiplication of the hologram $u_0[m_0,n_0]$ and the exponential terms in Eq.(3). 
\item Each GPU chip calculate the FFT of the result in Step 3 using CUFFT library \cite{cufft}. CUFFT library is the fast FFT library on the NVIDIA GPU.
\item Each GPU chip generate the propagation term $h[m_1-m_0,n_1-n_0]$ in Eq.(4).
\item Each GPU chip calculate the FFT of  the propagation term $h[m_1-m_0,n_1-n_0]$.
\item Each GPU chip  calculates the complex multiplication of the results of Steps 4 and 6.
\item Each GPU chips calculates the inverse FFT of Step 7.
\item Each GPU chips reduces the area of the results in Step 8 to $1,024 \times 1,024$ grids. 
\item Each GPU chips calculates the complex power (Eq.(5)) of the result in Step 9.
\item The CPU receives the four reconstructed images from the memories on each GPU.
\end{enumerate}

We used OpenMP library to operate each GPU chip because each GPU must be controlled by CPU threads. 
Each GPU can operate in parallel.
For real-time reconstruction, we repeat from Steps 1 to 11.

\section{Results}

\begin{figure}[htb]
\centerline{\includegraphics[width=12cm]{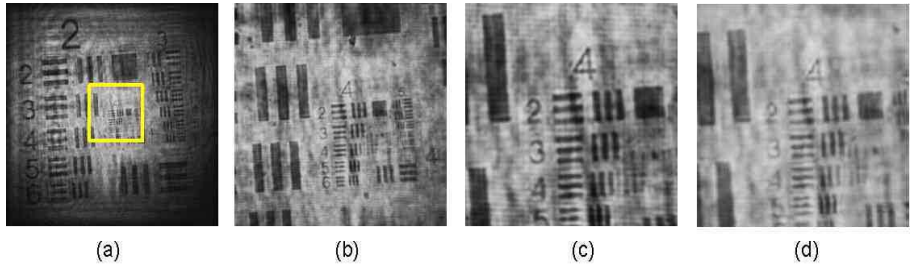}}
\caption{Reconstructed images using the Shifted-Fresnel diffraction. 
(a) Reconstructed image with $p_{x_1} \times p_{y_1} = 4.65 \mu m \times 4.65 \mu m$
(b) Reconstructed image with $p_{x_1} \times p_{y_1} = 4.65/2 \mu m \times 4.65/2 \mu m$
(c) Reconstructed image with $p_{x_1} \times p_{y_1} = 4.65/4 \mu m \times 4.65/4 \mu m$
}
\label{fig:img_shift}
\end{figure}

Figure 3 shows the reconstructed images of USAF 1951 test target using the Shifted-Fresnel diffraction. 
The Shifted-Fresnel diffraction can calculate a reconstructed image with different sampling spacings. 
 
Figure 3 (a) shows a large-area reconstruction with the sampling spacing $p_{x_1} \times p_{y_1} = 4.65 \mu m \times 4.65 \mu m$. 
Then, the area of the reconstructed image is about $4.8 mm \times 4.8 mm$. 
Figure 3 (b) shows a higher-resolution reconstruction in the yellow line of Fig. 3 (a) with the sampling spacing $p_{x_1} \times p_{y_1} = 4.65/2 \mu m \times 4.65/2 \mu m$. 
Figure 3 (c) also shows a higher-resolution reconstruction in the yellow line of Fig 3.(a) with the sampling spacing $p_{x_1} \times p_{y_1} = 4.65/4 \mu m \times 4.65/4 \mu m$. 
We can observe the reconstructed image in the minimum resolution of about $11 \mu m$.
  
Figure 3 (d) shows a reconstructed image using a bi-cubic interpolation algorithm which is an image enlargement method. 
In comparison with Fig.3(c) and Fig.3 (d), Fig.3(c) is a better quality image than Fig.3 (d).

Figure 4 shows an example of multiple reconstructed images from a hologram.
The DHM system can observe four reconstructed images simultaneously.

We used Intel Core2Quad Q6600 as the CPU, memory of 3Gbytes, and the operating system of Microsoft Windows XP Professional SP2. 
And, we used two GPU boards, NVIDIA GTX295, with the CUDA (Compute Unified Device Architecture) version 2.3 as a programming environment for the GPU chip.
The GPU board has two GPU chips on one, therefore, the DHM system can use four GPU chips.

\begin{figure}[htb]
\centerline{
\includegraphics[width=10cm]{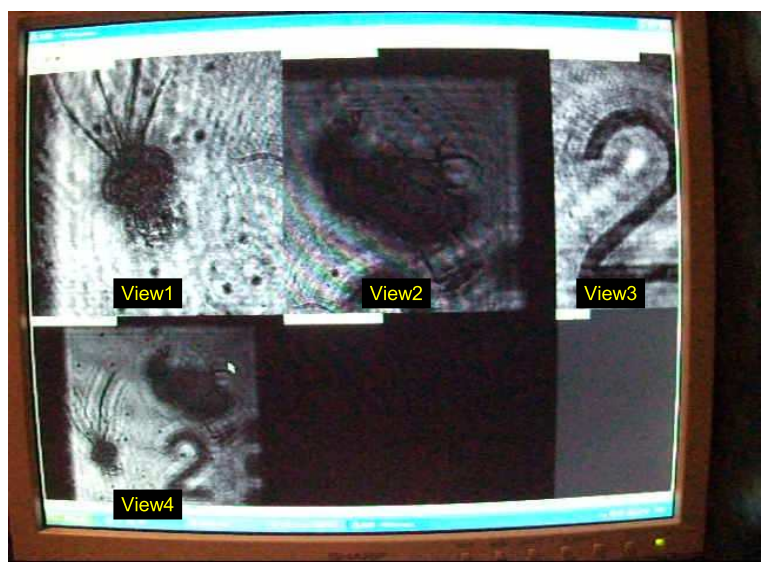}}
 \caption{ Multiple reconstructed images from a hologram using multi-GPU and Shifted-Fresnel diffraction (See the movie in Ref.\cite{movie}).}
\label{fig:img_multi}
\end{figure}

The hologram and four reconstructed images are $1,024 \times 1,024$ grids; however, in order to avoid circular convolution of Eq.(1) we expand the calculation area to double-size during the calculation. 
Under the condition that the number of multiple reconstructed images is four, the calculation time using the CPU alone is $9,542 ms$, whereas that using the four GPU chips is about $60 ms$.

In the movie of Fig.4, we can observe four reconstructed images in real-time (See the movie in Ref.\cite{movie}).
"View 1" shows  the reconstructed image of the head of mosquito.
"View 2" shows  the reconstructed image of the fly.
"View 3" shows  the reconstructed image of the USAF test target.
"View 4" shows  the reconstructed image with large viewing area of about $4.8 mm \times 4.8 mm$.
All of the views can be observed in arbitrary resolution, depth and positions.

\section{Conclusion}

In this paper, we presented the DHM system observable in multi-view and multi-resolution.
For multiple reconstruction, we used the four GPU chips.
In addition, we used the Shifted-Fresnel diffraction to obtain reconstructed images with arbitrary depth, resolution and view position. 
The method enables multi-view reconstructed images with the large area and higher resolution to be observed.

In future research, we will try a super-resolution method such as Ref.\cite{super}  in order to observe more resolution of a reconstructed image, and, we will develop the computational system using more multiple GPUs in order to obtain more reconstructed images at the same time.
  
This research was partially supported by the Ministry of Education, Science, Sports and Culture, Grant-in-Aid for Young Scientists (B), 21760308, 2009, and, the Ministry of Internal Affairs and Communications, Strategic Information and Communications R\&D Promotion Programme (SCOPE), 2009.

\end{document}